# Biomimetic Omnidirectional Anti-reflective Glass via Direct Ultrafast Laser Nanostructuring

*Antonis Papadopoulos, Evangelos Skoulas, Alexandros Mimidis, George Perrakis, George Kenanakis, George D.Tsibidis, and Emmanuel Stratakis*

A. Papadopoulos, E. Skoulas, A. Mimidis, G. Perrakis, Dr. G. Kenanakis, Dr. G. D.Tsibidis, Dr. E. Stratakis
Institute of Electronic Structure and Laser (IESL), Foundation for Research and Technology (FORTH), N. Plastira 100, Vassilika Vouton, 70013, Heraklion, Crete, Greece.
E-mail: stratak@iesl.forth.gr

A. Papadopoulos, E. Skoulas, A. Mimidis, G. Perrakis, Dr. E. Stratakis
Materials Science and Technology Department, University of Crete, Vassilika Vouton, 71003 Heraklion, Crete, Greece.

Keywords: ultrafast lasers, nano-structuring, biomimetic, anti-reflection, glass

We report on a single-step, biomimetic approach for the realization of omnidirectional transparent antireflective glass. In particular, it is shown that circularly polarized ultrashort laser pulses produce self-organized nanopillar structures on fused silica ($SiO_2$). The laser induced nanostructures are selectively textured on the glass surface in order to mimic the spatial randomness, pillar-like morphology, as well as the remarkable anti-reflection properties found on the wings of the glasswing butterfly, Greta oto[1] and various Cicada species[2]. The artificial structures exhibit impressive anti-reflective properties, both in the visible and infrared frequency range. Accordingly, the laser-processed glass surfaces show reflectivity smaller than 1% for various angles of incidence in the visible spectrum for S-P linearly polarized configurations. While, in the near infrared spectrum, the laser-textured glass shows higher transmittance compared to the pristine. It is envisaged that our current results will revolutionize the technology of anti-reflective transparent surfaces and impact numerous applications from glass displays to optoelectronic devices.

There are numerous species in nature that exhibit extraordinary surface functionalities including plant leafs [3,4], insects [1,2,4], reptiles [5], even elasmobranch fishes and marine life [4,6]. These remarkable properties help those species to survive, feed and thrive through extreme environmental conditions; their development has followed millions of years of evolution and the necessity of the species to evolve through natural selection. In most cases, the unconventional surface properties and attributes stem from their unique hierarchical morphological features, ranging from a few tens of nanometers to hundreds of micrometers in size [7].

Based on the concepts and underlying principles discovered by nature, an interdisciplinary field has been developed, aiming to design and fabricate biomimetic structures [7]. Research in this field indicated several methodologies to develop bioinspired surfaces, exhibiting hierarchical structuring at the nano- and micro- length scales [8–11]. Laser fabrication is a maskless process allowing material modifications with a high precision over size and the shape of the fabricated features [12]. However, due to optical diffraction, the feature size resolution is limited to the order of wavelength (i.e. microscale); therefore the challenge in biomimetic laser processing is to beat the diffraction limit and realize the structural complexity of natural surfaces, also at the nanoscale. Materials structuring using ultrashort (less than 1 picosecond (ps)) laser pulses, in particular, proved to be a precise and highly versatile tool to realize artificial surfaces that quantitatively mimic the morphological features and functionalities of their natural archetypes [13–19]. This capability comes as the outcome of the optimal combination of the ultrafast laser field and material properties that enables the production of features with sizes beyond the diffraction limit (i.e. nanoscale). A prominent example is the ultrafast laser induced formation of self-organised subwavelength periodic surface structures (LIPSS), which has been proven an important asset for the fabrication of nanostructures with a plethora of geometrical features[13,20–25].

This work is the first report on direct laser nanofabrication of biomimetic omnidirectional anti-reflective glass surfaces. It was inspired from the unique anti-reflection properties of the scales of the glasswing butterfly, *Greta Oto* and the cicada *Cretensis* species[2,26]. This property is due to the presence of arrays (with periodicity in the range of 150-250 nm) of nonreflective nanosized (sub-100nm size) pillars on both the top and the bottom surface of the scales. The current state of the art technologies employed for the production of anti-reflection surfaces require either complex multiple steps and time consuming procedures or chemical processes[8,27–33], which in some cases produce hazardous wastes. At the same time, the chemical coatings' quality tends to degrade with time [28,34,35]. Here, we demonstrate a single-step laser texturing approach for the structuring of biomimetic anti-reflective nanopillars, on Fused Silica glass ($SiO_2$) surfaces. The overall properties of the produced surfaces were found remarkably similar to the natural butterfly and cicada archetypes, both in terms of the surface morphology, as well of functionality. The proposed technique is simple, low cost, chemicals free and it can be easily scaled up using commercially available industrial laser processing systems [36,37].

Anti-reflective transparent materials can be useful for a wide range of technological applications including daily device screens[38], solar windows[39,40], optoelectronic devices[41,42] and a plethora of optical components [9,33,43,44]. We envisage that the laser-based biomimetic fabrication approach presented here will have a huge impact on such applications, considering the simplicity of the process coupled with the durability of the anti-reflective structures attained.

Following irradiation of Fused Silica with a specific number of linearly polarized pulses, NP, of appropriate fluence, Fl, high spatial frequency LIPSS (HSFL) are formed on the surface. The periodicity of HSFL is half, or less, of the laser wavelength λ, while their orientation is perpendicular to the incident polarization, as shown in **Figure 1a**. The physical mechanism that determines the size and orientation of HSFL in Fused Silica has been discussed in previews reports [21,22,45]. According to the most prominent explanation, the HSFL nanoripples' formation is the outcome of the interference between the incident field with the near-field scattered wave by localized high-electron-density hot spots [45].

HSFL can be considered as a suitable structure for the production of anti-reflection surfaces in the visible, as their characteristic periodicity, obtained by 2D-FFT analysis, is 359 ± 24 nm, i.e. within the range of the periodicity exhibited by the array of nanopillars formed on the cicada *Cretensis*[46] (**Figure 2a-2d**) scales.

Notably, irradiation of Fused Silica with a specific number of circularly polarized pulses at 1026nm, of appropriate fluence, gives rise to arrays of nanopillars (nanospikes), as shown by the respective SEM images presented in Figure 1d. The respective 2D-FFT image confirms the uniform distribution of such nanospikes within the spot area. The size of nanospikes is controllable by the wavelength of the incident beam, although the periodicity remains unaffected (supplementary **Figure S4**).

A parametric study was conducted to identify the appropriate laser parameters required for the nanopillars formation. It is generally observed that depending on the Fl used, nanospikes are formed at a different NP. For example at Fl=6.1 $\frac{J}{cm^2}$, well-ordered nanospikes are formed at NP=50, while at Fl=6.6 $\frac{J}{cm^2}$, well-ordered nanopillars are formed at NP=15, as shown in Figure 2f. By contrast, for NP > 50 no structures are produced while a crater is formed instead. Hence minor changes in fluence or in number of pulses are strongly influencing the formation of nanospikes. The range of fluences where nanospikes are formed is 5.7 – 8.3 $\frac{J}{cm^2}$, while that of number of pulses is 6-50 for the 1026nm central wavelength. Moreover, the optimum pulse duration for the formation of nanospikes is below 5ps. More information and the full map of structures obtained at various fluences and number of pulses can be found at the supplementary section (supplementary **Figure S2-S3**).

To account for the periodicity, radius of curvature and height of the fabricated nanospikes a morphological analysis has been performed. The periodicity, in particular, calculated from 2D-FFT analysis of the corresponding SEM images and presented in **Figure 2i and 2j,** is estimated to be within the range of 200-400 nm. It can be observed that the pseudoperiodic character of those structures gives rise to a wide range of periodicities. However, as the fluence and number of pulses increase, the mean periodicity slightly increases. The nanospikes radius, also calculated from the SEM images, shown in Figure 2l and 2m, was measured in the range of 70-100 nm. The radius values seem to be unaffected by the fluence changes, however they increase slightly upon increasing the number of pulses. Finally, the average height of nanospikes was estimated, from cross-sectional SEM images, to be 224±41 nm.

Towards the fabrication of large-area nanospike surfaces, a line scan study was initially attempted. It is observed that the homogeneous production of nanospikes, within the line, is quite challenging (see supplementary **Figure S7**). In order to avoid the textures' inhomogeneities, the nanospikes were progressively formed using a series of consecutive scans (see supplementary information **Figure S8**). For the formation of the nanospikes in large areas the sample was scanned in both directions and structuring is performed via the line-scan procedure described above. **Figure 3a**, presents a picture of a fused silica plate processed in its central part to provide a square-shaped area of nanospikes (SEM images are provided in **Figure S9**). It can be observed that under white light illumination, the laser-treated area is highly anti-reflective compared to the untreated outer peripheral area. The anti-reflection property of the processed areas has been confirmed by the respective reflectance spectra (Figure 3b and 3c) showing that it is more pronounced in the VIS/NIR part of the spectrum. It is also found that the scanning lines' separation significantly affects the anti-reflection effect (supplementary **Figure S10**). In particular, as the percentage coverage of nanospikes increases, the anti-reflection property is enhanced as well. Furthermore, as also shown in Figure 3b and 3c, the reflectivity was further decreased if both sides of the fused silica plate are laser-treated. Indicatively, the relative difference in reflectance between the untreated and the single-side processed plate is 1.7% for 1200nm and 1.9% for 600nm while that of the double-side processed one is 4.1% for 1200nm and 7.1% for 600nm.

Angle-resolved reflectance and transmittance spectra were recorded at three different wavelengths (445nm, 535nm and 658nm) for both *s* and *p* polarized light, using a custom-built angle-resolved reflectance spectroscope. The results are presented in **Figure 4a and 4b** respectively for the untreated, single- and double-side laser processed fused silica plate at various angles of incidence (AOI). It is observed that the reflectivity of single- and double-side processed fused silica plate is reduced compared to the bare one at practically all AOI and for all three wavelengths. Notably, the double-side processed material shows a significantly reduced reflectivity, which is less than 1%, for low AOI and less than 5% even at a large angle (i.e.60º), which reveals remarkable omnidirectional anti-reflective property. Furthermore, the transmission is higher than 85% in all cases and slightly decreases for large angles. This result indicates the omnidirectional high transparency of the laser-processed fused silica plates.

To analyze the anti-reflection properties of laser fabricated nanospikes, numerical simulations of light intensity distribution, as well as the angular dependence of reflection were carried out. The simulations were conducted on a 1.3 × 0.65 μm$^2$-area, 1mm-thick, supercell comprising, on both sides, silica nanospikes of four different mean-valued dimensions, deduced from SEM images, as well as on a 0.328 x 0.328 μm$^2$, 1mm-thick, unit cell comprising in both sides a single type of silica nanospike for comparison; on the other hand, reflection calculations from a 1 mm silica slab with a flat surface was used for reference.

In **Figure 5a, 5b and 5c**, the calculated against the measured reflectivities were compared, shown in Figure 4a and 4b, for the three different wavelengths and two polarizations. It is evident that regardless of the angle of incidence, the nanospikes' surface exhibits a decreased reflectivity by up to 10 times compared to an untreated glass plate. Notably, the observed anti-reflective behavior can be attributed to the presence of nanospikes that induces a progressive transition of the effective refractive index, allowing effective coupling of light into the bulk of fused silica. This light-coupling effect is further indicated by the significantly enhanced intensity of light among the nanospikes (Figure 5d). Supplementary **Figures S11, S12 and S13** confirm that nanospikes can reduce the reflection by almost an order of magnitude, compared with the plane silica surface, for all wavelengths within the range of 400-800 nm. Moreover, comparing supplementary **Figures S12** and **S13** at spectral/angular regions where severe diffraction does not occur (i.e. 500-800 nm and 0º-30º), the four-nanospike supercell shows a more broadband reflection reduction compared to the unit cell with the single-type of nanospike. This effect can be attributed to the nanospikes randomness, which is found also to be behind the unique omnidirectional anti-reflection properties of the glasswing butterfly [26].

In brief, we reported on a novel single-step and chemical-free technique for the fabrication of broadband, omnidirectional transparent anti-reflective surfaces, using ultrafast lasers. It is particularly showed that irradiation of fused silica with circularly polarized femtosecond pulses gives rise to the formation of subwavelength-sized nanospikes on the materials' surface. Notably, and unlike most anti-reflection surfaces fabricated to date, the laser induced nanospikes exhibit a quasiperiodic arrangement and present random height and width distributions, in a similar manner to the nanospikes found at the highly anti-reflective glasswing butterfly and Cicada wings. Our simulation analysis indicated that this random height distribution of nanospikes not only significantly reduces the surface reflectivity but also is the main reason for the observed broadband, omnidirectional anti-reflection property. Optical spectroscopy, indeed, confirmed that the reflectivity is suppressed by almost one order of magnitude over a large range of incident angles and wavelengths. Due to the nature of laser processing technology, the anti-reflective surfaces can be easily scaled up to large areas that could be used to enhance the light harvesting in solar cells or for efficient light emission in light-emitting diodes, as well as improving the performance of optoelectronic and electro-optical components and devices, including mirrors, lenses, photodetectors and displays [32,34,47].

**Experimental Section**

*Laser Processing:* Commercially available, UV-Grade polished samples of fused silica of 99.9% purity (purchased from GPO, Germany), with an average thickness of 1mm, were used. For the fabrication procedure, an Yb:KGW laser source emitting linearly polarized pulses with pulse duration of 170fs, 1 KHz repetition rate and 1026nm central wavelength, was employed. To produce circularly polarized (LC) pulses, a quarter-waveplate (QW) plate was used, where the optical axis was set to 45 degrees with respect to the original linear polarization. The experimental setup, shown in **Figure S14**, also includes a system of a half-waveplate (HW) and a linear polarizer (LP), used to adjust the laser power, a dichroic mirror (DM) to guide the beam, a concave lens (CL) of focal length $f$=150mm to focus the beam onto the sample and a CMOS camera to observe the sample surface. The sample was placed on a computer controlled 3-axis translational stage. The spot size was calculated to be 41 μm in diameter at $1/e^2$ using a CCD camera. All experiments were conducted in ambient air, at normal incidence and tight focusing conditions.

*Morphological and Optical Characterization:* The morphology of the laser-fabricated structures has been analyzed by a field-emission scanning electron microscopy (SEM) (JEOL JSM-7500F). The surface structures' periodicity was determined by a two-dimensional fast Fourier transform (2D-FFT) analysis of the corresponding SEM images using the Gwyddion software (supplementary section A). While the structures' radius and height were calculated from the SEM images via the ImageJ software. The reflectivity of the flat and the processed surface areas, was measured via a Bruker Vertex 70v FT-IR vacuum spectrometer, using a PIKE specular reflectance sample holder with a fixed 30-degree angle of incidence. In order to cover a spectral range of 22500 - 4000$cm^{-1}$ (444-2500 nm), two different sets of optics were used: (a) for 22500 - 8000$cm^{-1}$ (444-1250 nm), a Quartz beamsplitter and a room temperature Silicon diode detector, while (b) for 7500 - 4000$cm^{-1}$ (1333-2500 nm), a broad band KBr beamsplitter and a room temperature broad band triglycine sulfate (DTGS) detector were used. In any case, interferograms were collected at 4 $cm^{-1}$ resolution (8 scans), apodized with a Blackman-Harris function, and Fourier transformed with two levels of zero filling to yield spectra encoded at 2 $cm^{-1}$ intervals. Before scanning the samples, an Aluminum mirror (>90% average reflectivity) background measurement was recorded in vacuum. Finally, the transmittance and reflectance of the processed areas at various angles of incidence were measured at 445nm, 535nm, 648nm wavelengths, emitted by respective continuous wave (cw) laser sources.

**Supporting Information**

Supporting Information is available from the Wiley Online Library or from the author.

**Acknowledgements**

This work has been supported by the project LiNaBioFluid, funded by EU's H2020 framework program for research and innovation under Grant Agreement no. 665337 and from Nanoscience Foundries and Fine Analysis (NFFA)–Europe H2020-INFRAIA-2014-2015 (Grant agreement no. 654360). Funding is also acknowledged from the General Secretariat for Research and Technology (GSRT) and Hellenic Foundation for Research and Innovation (HFRI), no. 130229/I2. The authors acknowledge the technical support and software development by Mr. A. Lemonis (scilabs.gr & biomimetic.tech), also Mr. G. Aerakis & Dr. A. Trichas, Arthropod Collections Curator, NHMC for his useful insight concerning the Cicada Cretensis species identification.

E.St. conceived the idea and finished the manuscript. A.P. and E.Sk. have equal contribution in designing and performing the experiments and the characterization and wrote the first draft of the manuscript. A.M has performed experiments on other type of glasses. G.D.T., G.P. and G.K. have performed the theoretical modeling and simulations. All authors discussed the results and participated in writing the final manuscript.

**Competing interests**: Patent filed with No. PCT/GR2018/000010

Received: ((will be filled in by the editorial staff))
Revised: ((will be filled in by the editorial staff))
Published online: ((will be filled in by the editorial staff))References

[1]   R. H. Siddique, G. Gomard, H. Hölscher, *Nat. Commun.* **2015**, *6*, 6909.

[2]   J. Morikawa, M. Ryu, G. Seniutinas, A. Balčytis, K. Maximova, X. Wang, M. Zamengo, E. P. Ivanova, S. Juodkazis, *Langmuir* **2016**, *32*, 4698.

[3]   C. Neinhuis, W. Barthlott, *Ann. Bot.* **1997**, *79*, 667.

[4]   P. Vukusic, J. R. Sambles, *Nature* **2003**, *424*, 852.

[5]   P. Comanns, C. Effertz, F. Hischen, K. Staudt, W. Böhme, W. Baumgartner, *Beilstein J. Nanotechnol.* **2011**, *2*, 204.

[6]   L. Wen, J. C. Weaver, G. V. Lauder, *J. Exp. Biol.* **2014**, *217*, 1656.

[7]   E. I. Stratakis, V. Zorba, in *Nanotechnologies Life Sci.*, Wiley-VCH Verlag GmbH & Co. KGaA, Weinheim, Weinheim, Germany, **2012**.

[8]   Y. F. Huang, S. Chattopadhyay, Y. J. Jen, C. Y. Peng, T. A. Liu, Y. K. Hsu, C. L. Pan, H. C. Lo, C. H. Hsu, Y. H. Chang, C. S. Lee, K. H. Chen, L. C. Chen, *Nat. Nanotechnol.* **2007**, *2*, 770.

[9]   P. B. Clapham, M. C. Hutley, *Nature* **1973**, *244*, 281.

[10]  A. Rahman, A. Ashraf, H. Xin, X. Tong, P. Sutter, M. D. Eisaman, C. T. Black, *Nat. Commun.* **2015**, *6*, 5963.

[11]  H. Savin, P. Repo, G. von Gastrow, P. Ortega, E. Calle, M. Garín, R. Alcubilla, *Nat. Nanotechnol.* **2015**, *10*, 624.

[12]  E. Stratakis, H. Jeon, S. Koo, *MRS Bull.* **2016**, *41*, 993.

[13]  E. Skoulas, A. Manousaki, C. Fotakis, E. Stratakis, *Sci. Rep.* **2017**, *7*, 1.

[14]  U. Hermens, S. V. Kirner, C. Emonts, P. Comanns, E. Skoulas, A. Mimidis, H.11

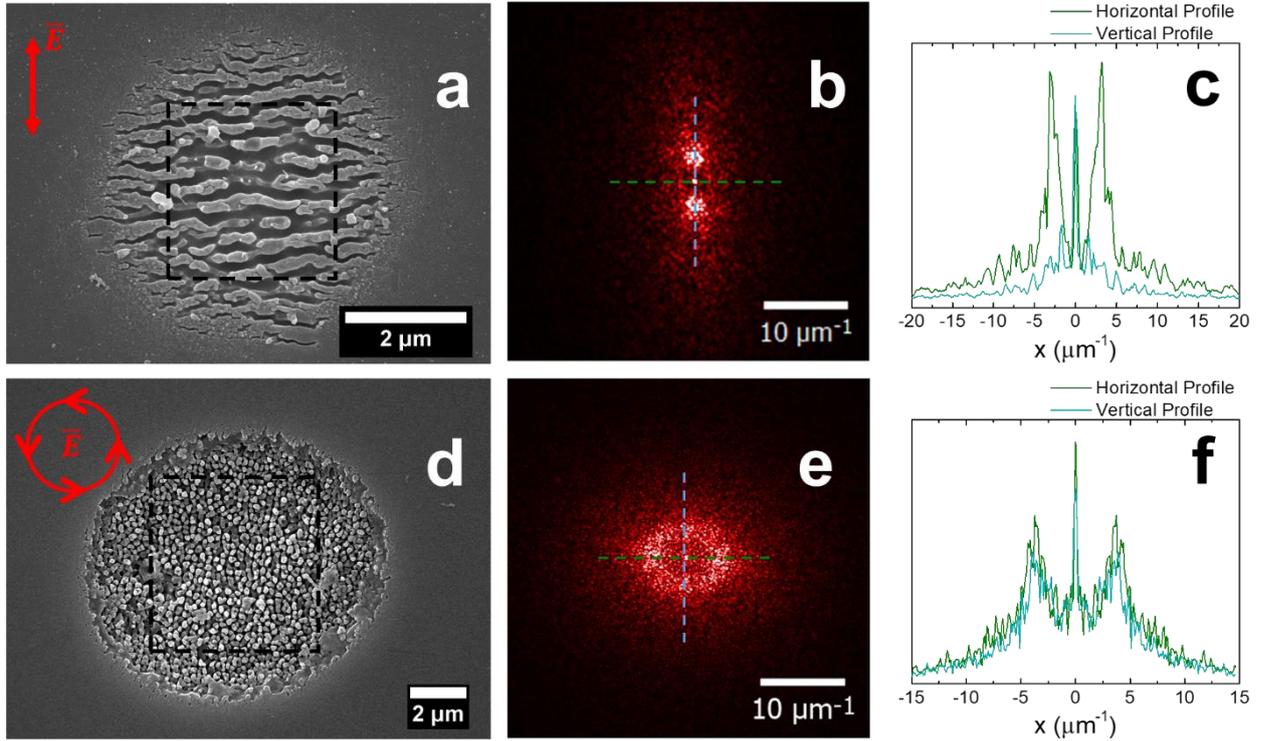

**Figure 1.** Polarization dependent LIPSS. **a)** and **d)** top-view SEM image of fused silica surface; **b)** and **e)** two-dimensional Fourier Transform (2D-FFT) of the area marked with the *dashed* black square; **c)** and **f)** the profile of the *dashed* horizontal and vertical line of Fourier Transform (b and e), following irradiation with **a-c)** NP=15 and λ=1026nm, linearly polarized pulses of Fl=2.8 $\frac{J}{cm^2}$, and **d-f)** NP=40 and λ=1026nm, circularly polarized pulses of Fl=6.0 $\frac{J}{cm^2}$. The red arrows (in the Left Column) indicate the laser beam polarization.

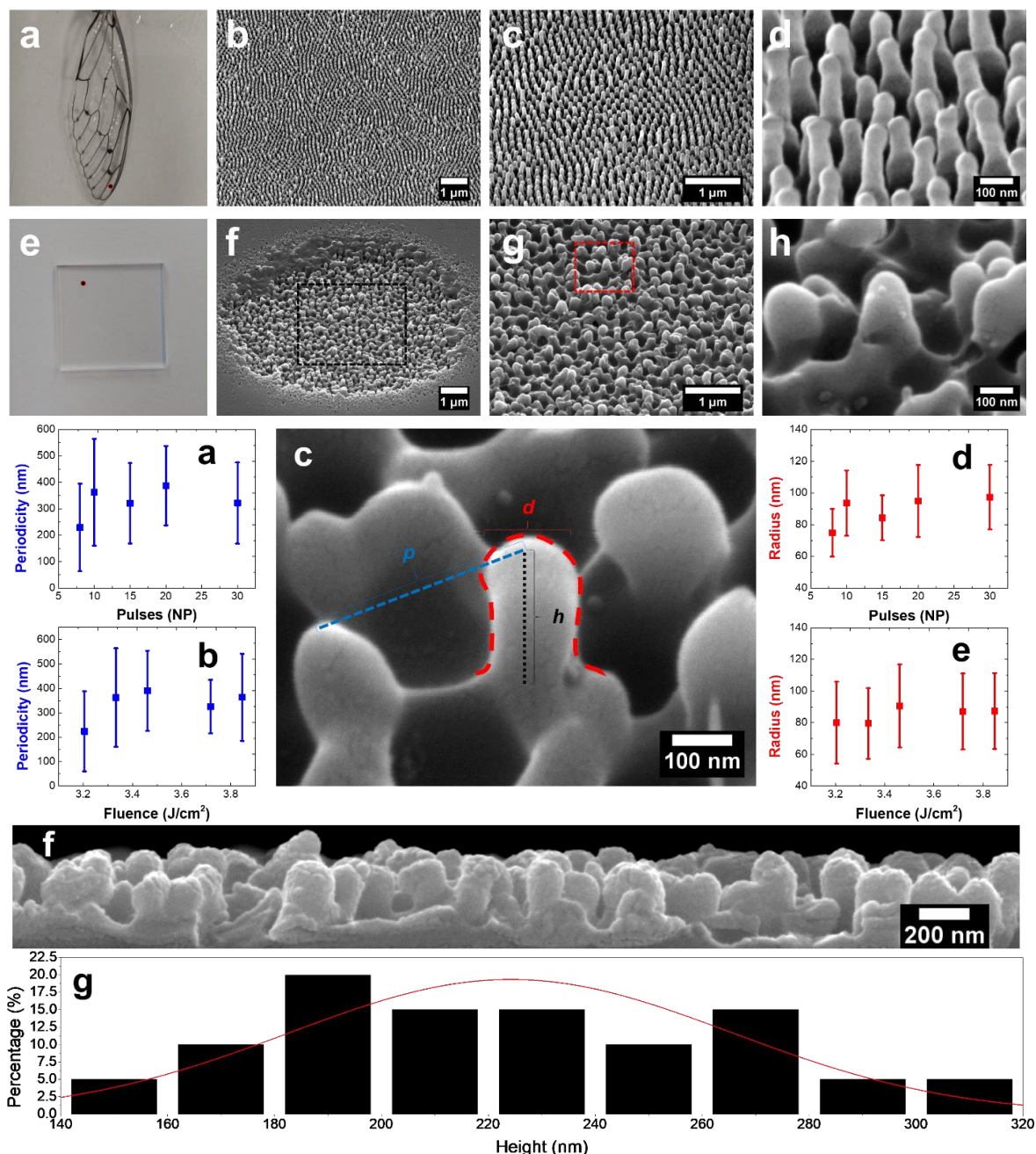

**Figure 2.** Natural *vs* biomimetic artificial surface and morphological characterization of the laser fabricated nanospikes; **a-d)** Photograph of a Cicada Cretensis[46] scale and respective SEM images (45º tilted) of the transparent anti-reflective area at different magnifications; the red spot indicates the SEM imaging area; **e-h)** Photograph of a fused silica plate and SEM images (45º tilted) of a spot fabricated on the surface using NP=15 circularly polarized laser pulses of λ=1026nm wavelength, 1kHz repetition rate with and170 fs pulse duration, of Fl=6.8 $\frac{J}{cm^2}$ ; the red spot indicates the location of irradiation; **i)** Nanospikes' periodicity as a function of NP for Fl=6.6 $\frac{J}{cm^2}$; **j)** Nanospikes' periodicity as a function of Fluence for NP=10; **k)** High resolution SEM image (45º tilted) of a single nanospike; **l)** Nanospikes' radius as a function of NP for Fl=6.6 $\frac{J}{cm^2}$ ; **m)** Nanospikes' radius as a function of fluence for NP=10; **n)** Cross-sectional SEM image of the fs laser induced nanospikes; **o)** Height distribution.

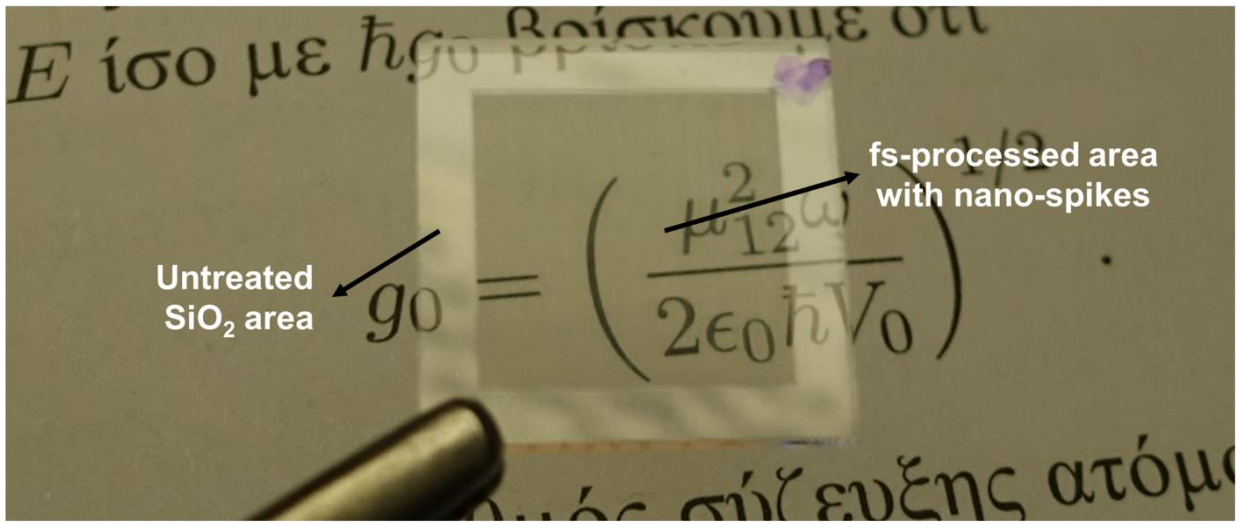
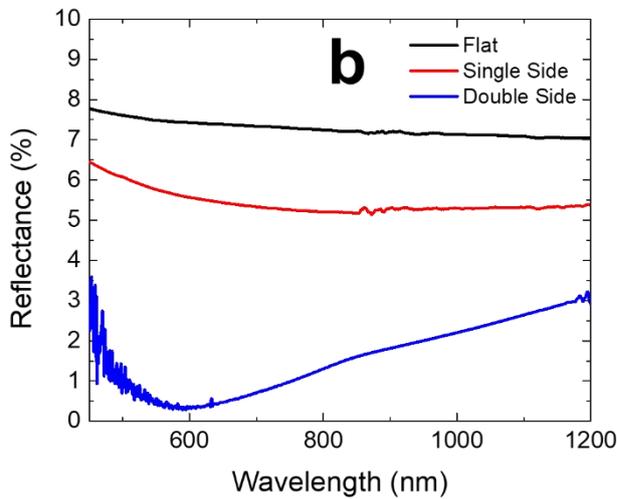
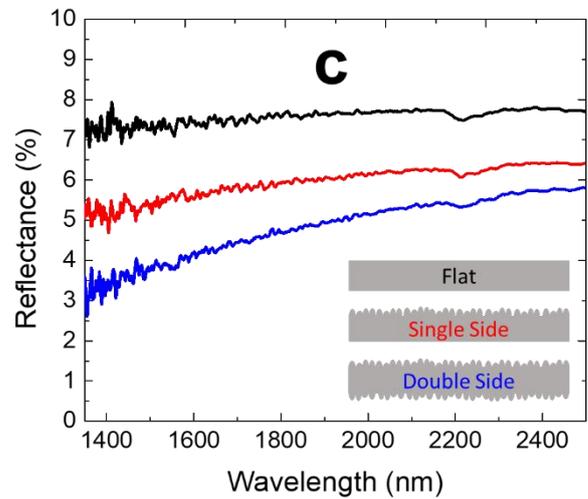

**Figure 3. a)** Photograph of a fused silica sample plate (background is from the Greek version of the book Quantum Optics, Mark Fox), the central part of which was laser-treated to fabricate nanospikes; the black dashed rectangle indicates the processed area. **b-c)** Reflectance spectra of a pristine (black lines) and laser-treated at one (red lines) or both sides (blue lines) fused silica plate. Laser processing has been performed using $N_{eff,\ 2D}=15$ circularly polarized pulses ($\lambda$=1026nm, 1kHz, 170fs) with Fl=5.8 $\frac{J}{cm^2}$.

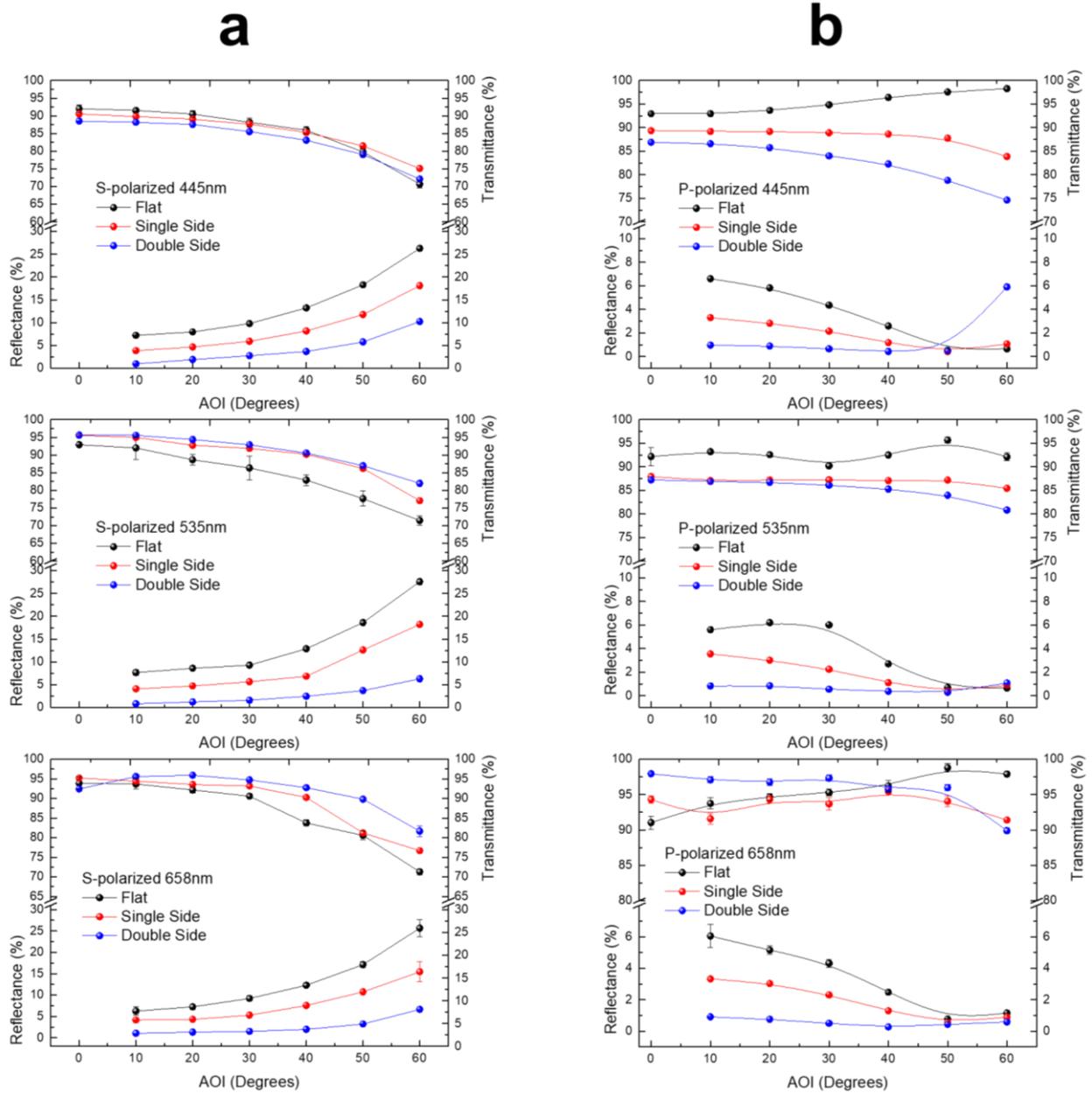

**Figure 4.** Transmittance and Reflectance measurements for various AOI at 445nm, 535nm and 658 nm wavelength. Transmittance and Reflectance measurements of *Flat*, *Single Side processed* and *Double Side processed* Fused silica for **a)** **s**-polarization and **b)** p-polarization configuration.

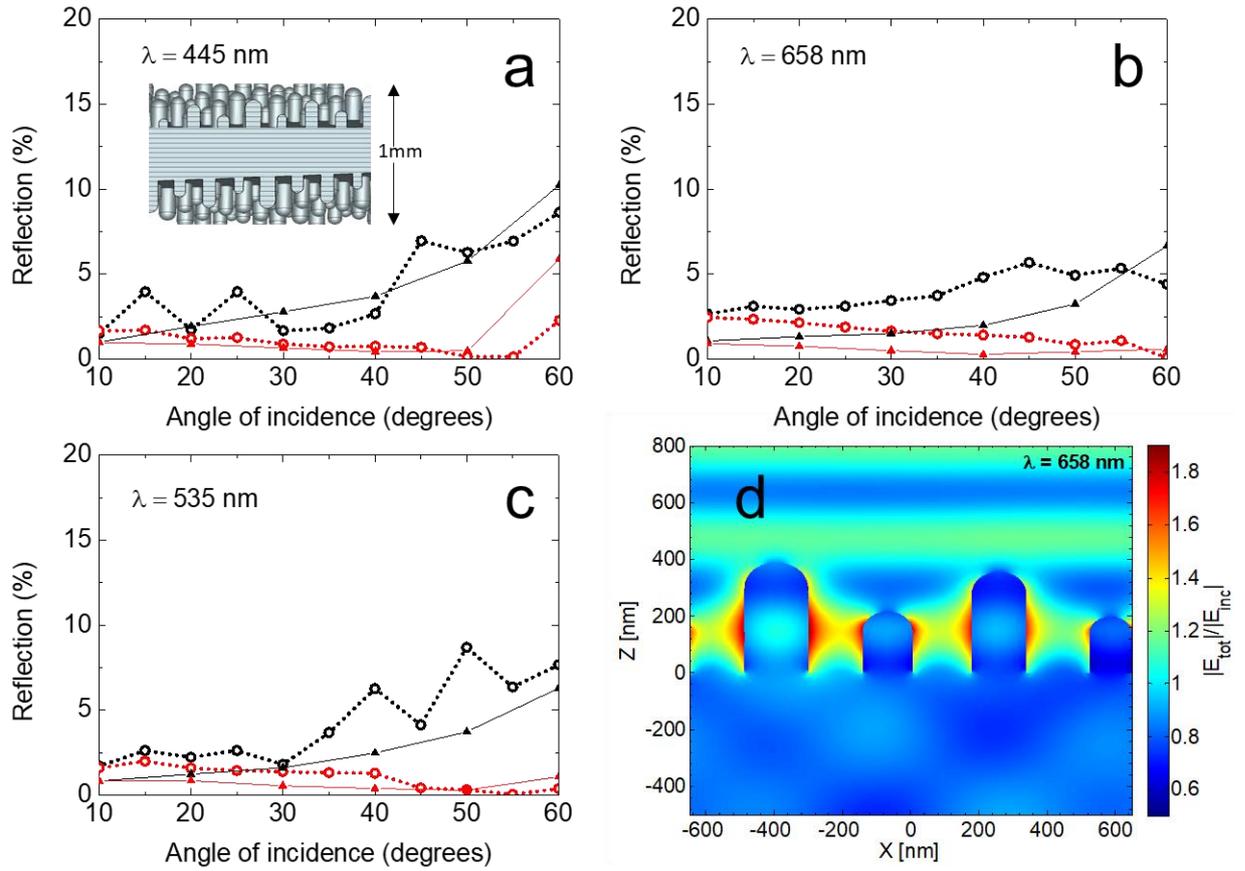

**Figure 5.** Theory (dashed line) vs experiment (continuous line). Reflectance measurements for various AOI at 445nm **(a)**, 658nm **(b)** and 535 nm **(c)** wavelength of Single Side processed (Black) and Double Side processed (Red) Fused silica. **d)** Distribution of Electric field $|\vec{E}|$ at wavelength 658 nm at normal incidence